# Model Optimization for A Dynamic Rail Transport System on an Asymmetric Multi-Core System

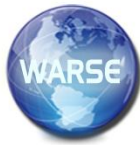

**Anas M. Al-Oraiqat**[*1], **Alexander Y. Ivanov**[2], **Yuriy A. Ivanov**[2]
[*1]Onaizah Colleges, College of Engineering & Information Technology,
Department of Cyber Security
Onaizah, Kingdom of Saudi Arabia, P.O. Box 5371
anasoraiqat@oc.edu.sa
[2]Donetsk National Technical University, Department of Computer Engineering, Ukraine
alex.ivanov.35a@gmail.com, e-mail: yuriy.o.ivanov@gmail.com

**ABSTRACT**

The problem of optimization of the rolling dynamics model is considered. That providing safe movement at high frequency when interacting with the railway. Moreover, allowing to evaluate the dynamic parameters when designing new and modernizing existing locomotives. The object of this research is a rail transport dynamic system model. The article subject is the optimization methods of the real-time software model. The article's purpose is to increase the efficiency of the digital hardware in the rolling stock loop model by optimizing the organization of the computing process. Furthermore, we take into account the different-frequency parameters of the model. Also, based on the principles of design the microarchitecture of modern multicore systems.

The mathematical model analysis of the object made it possible to attribute it to the class of hard real-time systems. The computation of the model phase variables with different frequencies is necessary to optimize the simulation time of the train movements and is performed by splitting the original algorithm into parallel threads. The NP-problem of nonlinear integer schedule optimization, which is solved by the metaheuristic algorithm, is posed. The developed planning algorithm and the cyclic schedule implementation for the model of a dynamic real-time object consider microarchitecture solutions of symmetric multiprocessor systems with shared memory and methods for optimizing software tools.

The experiments confirmed the operability of the optimized model. Also, allow us to recommend it for practical use in studying objects and determine the dynamic force of trolley structural elements during operation. These processes are necessary for the optimal choice of the scheme and rolling stock equipment parameters, as well as to reduce the dynamic forces acting on the structural elements of the train and the rail track. Prospects for further research may consist of optimizing the distribution of program threads by the criterion of connectivity of the model phase variables, as well as an experimental study of the proposed methods while expanding the class of problems being solved.

Analysis of the optimized model simulation results, using cyclic schedules shows the correspondence of the obtained simulation results to the standard. The main advantage of the model is the increase in productivity when performing data processing by reducing the processor time. The optimized cyclic schedule algorithm of the semi-natural modeling platform is used for the subsequent development of the control system in real and accelerated time scales.

## 1. INTRODUCTION

The main tasks of the rail vehicle dynamics are to study the processes of oscillations caused by the interaction of the railway trolley (cars, locomotives) and tracks [1]. When solving them, the Hardware/Software in the Loop (HIL) technology is used, which allows testing control algorithms together with the control object model in the Real-Time(RT) mode at the initial stages of design [2]. Besides, software/hardware simulation provides constant verification of the model-oriented process design, including technical specifications, modeling, and rapid prototyping[3] of the elements' optimal parameters of the control system [4].

The RT model under consideration belongs to the class of models whose mathematical description contains systems of stiff Differential Equations (DEs) [5]. The phase variables frequencies of hard RT models vary over a wide range. Providing such characteristics requires a special organization of computations when creating RT models [6].

To increase the efficiency of controlling the process of modeling the dynamics of rail vehicles, high-performance multi-core Chip Multi-Core (CMC) processors, asymmetric multi-core (AMC), and distributed simulators are used. When designing a parallel multi-threaded model, the hardware platform, the data model corresponding to the code architecture and memory organization are taken into account [7].The microarchitecture of modern multi-core chips supports hardware shared memory. In a system with shared memory, system performance is determined by the cache hierarchy, which allows it to base on the principles of data-oriented design [8]. Task scheduling policies used in modern concurrent programming environments may not improve concurrent application performance or provide RT mode. The optimal planning of a hard dynamic system model for the CMC architecture is based on solving the NP-task of integer programming. When preparing an information model, it is enough to use the static load balancing method to distribute the computational load for the system integrating parts of Ordinary Differential Equations (ODE) between processors.

The article's goal is to optimize the software model of the RT dynamic system in the architecture of a standard CMC-based system. Whither all cores of the parallel architecture have the same performance.

The second section discusses the features of the mathematical model of the HIL rail transport system. In the third section, the analysis of known tools for studying the model is carried out, the criterion for the existence of real-time, the choice of the structure of parallel hardware is substantiated, and the problems encountered during optimization are indicated. Section 4 proposes a method for implementing multithreading during parallelization using flow blocks and optimization of the main stages of the organization of computations in the model. Section 5 presents the results of experiments into consideration, the microarchitecture features of modern computers. Section 6 is devoted to conclusions.

## 2. TASK FORMULATION

When moving, the locomotives make complex linear and angular movements. The study is divided into vertical, transverse horizontal movements and longitudinal vibrations in the train [9].The locomotive's mechanical model, the design scheme that is shown in Figure 1, is considered as a three-solid system.

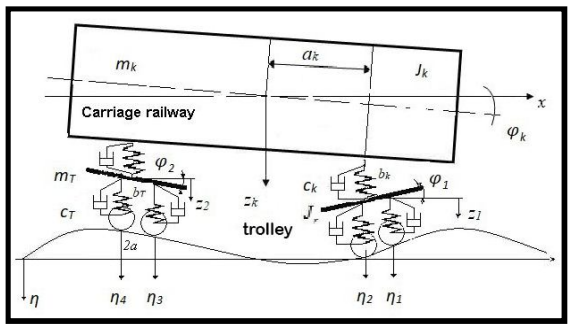

**Figure 1:** The wagon (locomotive) mechanical model in the ZX plane

Accepted Assumptions:
1. The carriage mass centers and the trolley coincide with their geometric centers.
2. Irregularities of both rail tracks are assumed to be the same.
3. Stiffness values and damping factors are the same for different wheelsets.
4. Elastic and dissipative forces act along the axis of the corresponding element.
5. The spring and the hydraulic vibration damper parallel to it are fixed at one point.
6. The wheelset and the mass of the path interacting with it do not take into account the stiffness of the wheel-rail contact and move continuously.

As a perturbation in the study of the model, geometric rails of the rails are taken. For the generalized coordinates of the model area translational movement of the entire system along the axis of the path, vertical movements of the carriage $z_k$, trolleys $z_1, z_2$, carriage turning angles $\varphi_k$, and trolleys $\varphi_T$ ($\varphi_1$ and $\varphi_2$).

To describe the vertical vibrations of a locomotive with springs and dampers, a mathematical model is used that includes 38 second-order DEs [10].The system linearization leads to the main phase coordinates, and its order decreases to the second. The vertical oscillations motion's equations of a locomotive with two-tier suspension are described by the following DEs [11].

Carriage vibration equations:
$$m_k * z_k'' + b_k * (2 * z_k' - z_1' - z_2') + c_k * 2 * (z_k - z_1 - z_2) = 0 \tag{1}$$

$$J_k * \varphi_k'' + a_k * b_c * (2 * a_k * \varphi_k' - z_1' + z_2') + a_k * c_k * (2 * a_k * \varphi_k - z_1 + z_2) = 0 \tag{2}$$

Oscillation equations of the first trolley:
$$m_T * z_1'' - b_T * (z_k' - z_1' + a_k * \varphi_k') - c_k * (z_k - z_1 + a_k * \varphi_k) + 2 * b_T * z_1' + 2 * c_T * z_1 = b_T * (\eta_1' + \eta_2') + c_T * (\eta_1 + \eta_2) \tag{3}$$

$$J_T \cdot \varphi_1'' + 2 \cdot a_T^2 \cdot b_T \cdot \varphi_1' + 2 \cdot a_T^2 \cdot c_T \cdot \varphi_1 = a_T [b_T (\eta_1' - \eta_2') + c_T (\eta_1 - \eta_2)] \tag{4}$$

Oscillation equations of the second trolley:
$$m_T * z_2'' - b_T * (z_k' - z_2' + a_k * \varphi_k') - c_K * (z_K - z_2 + a_K * \varphi_K) + 2 * b_T * z_2' + 2 * c_T * z_2 = b_T * (\eta_3' + \eta_4') + c_T * (\eta_3 + \eta_4) \tag{5}$$

$$J_T \cdot \varphi_2'' + 2 \cdot a_T^2 \cdot b_T \cdot \varphi_1' + 2 \cdot a_T^2 \cdot c_T \cdot \varphi_2 = a_T [b_T (\eta_3' - \eta_4') + c_T (\eta_3 - \eta_4)] \tag{6}$$

with the following notations and their meanings:
$m_k$ = 57tcarriage weight, $J_k$ = 70 carriage moment of

inertia, $m_T$ = 9ttrolley weight, $a_T$= 3,725 m half carriage base, $c_T$ = 3040 kN/m, $b_T$ = 30 kN*s/m stiffness and damping in the first tier (trolley), $c_k$ = 2660 kN/m, $b_k$ = 100 kN*s/m stiffness.And damping in the second tier (carriage),$z_i$, $z_i'$, $z_i''$, $\varphi_k$, $\varphi_k'$, $\varphi_r''$ generalized coordinates and their derivatives concerning time, $\eta_j(t)$indignation from the side of the path under $j$ wheelset. The external kinematic disturbance of the system is unevenness on the side of the path.As a perturbation, we used the averaged geometric model of roughness [11], described by the following equation:

$$\eta(t) = A1\sin(wt) + A2\sin(2wt) \quad (7)$$

where$A1$and$A2$the amplitudes of the first and second harmonics of the sinusoidal irregularities along the length of the rail link, and are selected depending on the type and condition of the track.Roughnesses of rails for the system with a transport delay$\tau = 2a_k/v$ between trolleys is determined:

$w = 2\pi V / L$
$n_1(t) = n_0(\sin wt)$
$n_2(t) = n_1(t)(t - 2a_1/V)$
$n_3(t) = n_1(t)(t - 2a_2/V)$
$n_4(t) = n_1(t)(t - 2(a_1 + a_2)/V) \quad (8)$

$n_{11}(t) = n_0 w(\cos wt)$
$n_{22}(t) = n_1(t)(t - 2a_1/V)$
$n_{33}(t) = n_{11}(t)(t - 2a_2/V)$
$n_{44}(t) = n_{11}(t)(t - 2(a_1 + a_2)/V).$

The displacement $\eta_i$ and acceleration $\eta_i'$ are determined with the delay at the moments $2a_1/V$, $2a_2/V$, $2(a_1 + a_2)/V$ for the corresponding wheels, where $a_1$ =0.005m and $a_2$ =0.002m are the amplitudes of flatness.

The solution of a simplified system of DEs can be obtained analytically. However, given the subsequent development and addition of the model with the introduction of nonlinear functions, more efficiently are performed by modeling. An analysis of the solutions shows that the natural oscillations of the masses of the wagon and the trolley are the sum of two harmonic oscillations: high-frequency mass oscillations on a rail base and carriage vibrations on springs. The second type frequencies of oscillation vary in the range of 0.2-2.6 hertz and significantly lower than the frequencies of the first type of oscillation 11.2-21.7 hertz [12]depending on the weight of the locomotive, the load weight, carriage type, and the stiffness of the spring set of the carriage. Thus, the presented object is multi-inertial with different directive periods for computing the right-hand sides of the equations of the RT model. Moreover, for this object the proposed optimization methods are relevant.

The execution time optimization of the object parallel program model under consideration should include the steps:
1. Choosing a hardware structure with a modern microarchitecture for developing a data model.
2. Organization of the computational process of a system model numerical integration and resource allocation.
3. Creation and research of a simulation program into consideration, the data decomposition, and operating system features.

## 3. RELATED WORKS

The standard for software verification of rail locomotive models is currently the Manchester test [11]. Methodology for evaluating mathematical models USDOTVolpe National Transportation Systems Center [13].One of the most functional and widely used in the industry is the ADAMS/Rail system [14].The closest tasks to be solved for studying longitudinal vibrations of locomotives are works [15, 16].This work, in addition to optimization, is distinguished by into consideration, the locomotive's vibrations concerning the *y*-axis (galloping), which affects the ergonomic characteristics of the train.

A necessary condition for the existence of a cyclic schedule RT:$\sum_{j=1}^{N}(\tau_j/T_j) \leq 1$ [17],set policy terms $T_j$and computation times$\tau_j$for *N* phase variables, determines the choice of the modeling system architecture.If the condition is not fulfilled or complication (increase$\tau_j$) of the model is expected in the future, it is advisable to use the parallelization of the program algorithm. For models with long policy terms$T_j$and small$\tau_j$RT can also be provided on a single-processor system.

The choice of a universal CMC for modeling is most consistent with the structure of the HIL system [18], in which digital processors behold the general memory address space and exchange variables on a general bus. Also, support for such a subclass of SM-MIMD systems, in the Flynn classification, is available in leading operating systems. Into consideration, the different inertia in the model is based on hardware [19] and the special organization of the computing process. Computations are planned by static cyclic time schedules[20]that satisfy RT conditions [21].

Organization of the computational process of a system model numerical integration and resource allocation for CMC is a mixed problem of non-linear integer programming of NP- difficulties.To solve it, it is advisable to apply the methodology of metaheuristic algorithms[20]and, in particular, an artificial bee colony [22].Along with algorithms (numerical integration) and thread processing methods in parallel programming, it is accepted [23]. The speed of high-performance applications will be higher if all available cores of modern processor systems are used and the main memory requests are executed in the nearest cache. In this case, the model developer must take into allow the principles of parallel programming and use one or several methods of the OpenMP, MPI, POSIX, or Win32 API threads.

Using well-known low-level optimization methods: compiling a program by an optimizing compiler, system measurements, and profiling a program also improves the overall performance of the model. Should be considered, the modern programming systems for parallelism can create models with unpredictable behavior due to data races and locks [24]. The tasks solution with considering the analysis of the related works will improve the efficiency of the updated parallel model.

## 4. METHODS

Multithreading implementation methods used by Intel [25] include analysis, development, implementation, and program specialization. The evaluation criterion for parallel RT software models is runtime. During development, the program model is described in terms of parallel control models and data-parallel utilization, oriented to scientific applications may differ from general-purpose architectures.

### 4.1. A Mathematical Model for Optimizing a Computing Process.

Each RT model is unique and requires a sound implementation of the modeling process. The main influence on the organization of computations of the program model is exerted by the particularities of solving systems of hard ODEs that describe the object of study. The HIL model of DEs [5] cyclically operates: entering information into the digital part of the system, computing each $i^{th}$ phase variable for integration over time $\tau_i$, and outputting computed values to the hardware of HIL models. The phase variables of a hard control system have different frequencies. Also, they must be cyclically computed with periods $T_i$ by parallel program threads. This requires controlling the integration of $N$ phase variables as a static cyclic schedule with differing guide dates [26] at each integration step. In models of important dynamic systems that have been developed and used for a long time, information on the intervals of work is known in advance. The choice of parallel threads for model development is explained as follows: threads of one process are created with their context and have common resources. Switching between threads is quick compared to task switching, which takes about a thousand processor cycles.

DE models are transformed into a system of first-order equations. For numerical integration of phase variables in threads, the Euler method is chosen. This minimizes the computation time of the right-hand sides of the DEs system and the memory amount in the inline substitution case. Instructions going beyond the cache limits of the first level L1i causes a significant decrease in performance, much more than when processing data. For processors, the execution time of instructions when leaving the L1i cache is increased three times [27].It should also be noted, that the second level L2 cache stores both code and data that can claim the same cache lines, as a result of the performance will drop when the L2 size is exceeded. Thus, the proposed choice solves the problem of algorithmic optimization.

The planning of numerical integration threads is carried out based on the Earliest Deadline First (EDF)and Proportional Fair (PF) approach [28]. The purpose is the proportional parallel execution of all task threads during the base period. In this case, the cyclic execution of the computation thread of each phase variable over the entire integration interval occurs with a constant frequency. The Easy Release (ER) effect planning is achieved by dividing the program threads into blocks. The proportional execution idea of parallel flow blocks at the integration step of the DE system is used in this work. For equable execution of the entire thread, blocks should be executed during time intervals of approximately equal length. To control the computational threads of the dynamic system, model Synchronous Data Flow (SDF) is used.

In the general case, for a multi-core system, the problem arises from the program blocks' optimal distribution of DE integration threads into cores while providing RT, which is a nonlinear integer optimization problem. Let the executive part of the modeling system be characterized by the following conditions: the computer performs $M$ threads with a period $T_i (i = 1,2, \ldots, M)$following the time sequence. The processing periods values of the threads can be sorted in ascending order: $T_1 < T_2 < \ldots < T_m$. Each thread needs processor time $\tau_i (i = 1,2, \ldots, M)$, and each thread must be executed before the next request arrives. All threads must be executed once until the next thread execution cycle arrives after the full period$T_c = leastcommonmultiple\ (T_i), (T_c \in N)$. Let $L\ (L \in N)$be the duration of the cycle RT, during which a certain fraction $\Delta_i\ (0 < \Delta_i \leq 1)$ of each thread is executed.Thus, the execution of each thread will synchronize with the value $L$. The window for executing the block of each thread is defined as $\Delta_i \tau_i$.In this case, the execution of the thread will be completed in $k_i = \left]\frac{T_i}{L}\right[$, RT-cycles, where$]x[$ is the operation of the largest integer, which is less than or equal to $x$. With the introduction of the RT - cycle, the system replaces the periods of the threads with values of $T_i'$, multiples of $L$ and not exceeding $T_i$: $T_i' = k_i L\ \leq T_i$.As a result, the value of the full period $T_c' = leastcommonmultiple\ (T_i')$ will also change. Then in the complete cycle, there will be $k = \left]\frac{T_c}{L}\right[$ RT-cycles.

The computation scheme implementation in the CMC system is complicated by the distribution of threads across specific cores. The mathematical structure analysis of this problem allows us to attribute it to the problem of combinatorial optimization [29].

To schedule the computational process of processing flow blocks for a multicore system, a parameter that displays the splitting of threads into groups is introduced [30].

Let be 0 if $i$ thread is not executed on the $j^{th}$ core:

$$x_{i,j} = \begin{cases} 0, if\ i\ thread\ is\ not\ executed\ on\ j-core, \\ 1, if\ executed\ processed. \end{cases}$$

Based on the necessary condition for the existence of a cyclic schedule RT, the load on each core cannot exceed 100%:

$$(\forall j) \sum_{i=1}^{M} \frac{\tau_{i,j}}{T_i} x_{i,j} \leq 1 \quad (10)$$

Moreover, each thread can run on only one core:

$$(\forall i) \sum_{j=1}^{n} x_{i,j} = 1 \quad (11)$$

When constructing a schedule, all $M$ modeling threads are considered:

$$\sum_{j=1}^{n} \sum_{i=1}^{M} x_{i,j} = M \quad (12)$$

Each RT-cycle in such a model has its period $L_j$, which is a natural number and cannot exceed the value of the minimum period of the processed threads $T_j^{min}$ min on the j-core of the processor. The effectiveness of the entire system, as a solver of the simulation task, is the sum of its RT-cycles and is determined by the following function:

$$F = \sum_{j=1}^{n} \left[ \sum_{i=1}^{M} \langle \left( \frac{\tau_{i,j}}{T_i'} - \frac{\tau_{i,j}}{T_i} \right) x_{i,j} \rangle + \frac{\sum_{k=1}^{M} x_{k,j} p}{L_j} \right] (13)$$

where $(\forall j, i) T_i' = T_i - \left\lfloor \frac{T_i}{L_j} \right\rfloor$

The access time to the system memory at this stage is not reviewed in the model, considering it to be a constant. Thus, a model for optimizing a computational process, that processes $M$ threads of a simulation task on $n$ cores, can be written as follows:

$$\arg \min F(L_j, x_{i,j})$$
$$(\forall j) L_j \in N, L_j \leq T_j^{min}$$
$$(\forall j, i) x_{i,j} \in Z, 0 \leq x_{i,j} \leq 1$$

$$F = \sum_{j=1}^{n} \left[ \sum_{i=1}^{M} \langle \left( \frac{\tau_{i,j}}{T_i'} - \frac{\tau_{i,j}}{T_i} \right) x_{i,j} \rangle + \frac{\sum_{k=1}^{M} x_{k,j} p}{L_j} \right] \quad (14)$$

$$(\forall j) \sum_{i=1}^{M} \frac{\tau_{i,j}}{T_i} x_{i,j} \leq 1$$

$$(\forall i) \sum_{j=1}^{n} x_{i,j} = 1$$

$$\sum_{j=1}^{n} \sum_{i=1}^{M} x_{i,j} = M$$

$$(\forall j, i) T_i' = T_i - \left\lfloor \frac{T_i}{L_j} \right\rfloor,$$

where $i = \overline{1, M}.\quad j = \overline{1, n}.$

The type of the proposed combinatorial model is determined by the objective function $F$, which depends on $L_j$ and $x_{i,j}$. The first part of the sum $F$ determines the increment of the processor usage due to a change of the base periods of the threads, when the RT cycle is introduced. The computation of $T_i'$ is a nonlinear function, as well as a graph (hyperbole) of the change in the sum second part of the objective function, that reflects a decrease in CPU time for switching between threads with increasing $L_j$. In addition, the objective function of the computation model for the CMC system includes Boolean variables $\bar{x}$. The graph of the function $F$ is a set of n-dimensional nonlinear discrete planes, that number is equal to the number of cores. Note that the overall performance indicator will correspond not to the minimum total of expenses on each core, but the sum of the minimum. This is since different values of the RT cycle period on each core are selected. At the same time, there is no accumulation of expenses introduced by changing the period when using the parameter $L_j$.

The main features of this optimization task are multi-criteria and multi-extreme. The optimized function and the range of its admissible values do not change in time. So that it can be assigned to the class of statistical combinatorial optimization problems. The problem of nonlinear integer optimization is solved using the artificial bee colony (ABC) algorithm [31], designed to find global extrema of complex multidimensional functions.

The proposed parallelization of control using flow blocks and the proposed organization optimization of computations in the model, have the greatest effect on increasing the productivity of the model. The parallelization procedure is difficult to automate and therefore, for complex and important dynamic systems, the best results are achieved by manual programming according to the logical structure of the model.

**4.2. The Allocation of the Model System for the Implementation.**

Design solutions for creating a model should take into account the implementation features of multithreading on a hardware platform. The hardware scheduler of the processor core queues microoperations on the corresponding port for execution by processor modules [32]. With Hyper-Threading (HT) technology, threads from one core can simultaneously perform several operations, but cannot simultaneously save data from the core to the L1d data cache via the Store Data port. This disadvantage is absent in the separation of threads between different physical cores. Further, when running threads on the core, buffered writes can over thread core write buffers. It will block the cache until the data upload that caused the cache miss is completed [33].To ensure the equable execution of threads from both inputs, the HT core tries to balance the values of its counters. Which tracks the number of clock cycles of code execution from each of the inputs. Additionally, one of the reasons for switching threads on the core is missing the necessary data in the cache. While the data will be read into the cache from the next memory level, the core switches the pipeline and execution devices to another input. If during the execution of the thread from the second input a cache miss also occurs, then the core will switch back to the first input. There may be a case where the core will constantly switch between two inputs. In this case, for each thread, it is required to reload the cache filled with the previous thread, i.e. data in the cache will be reloaded all the time. This applies to both the L1 cache and the L2 cache, as it will take longer to reload.

Thus, when developing a model of a hard RT dynamic system to reduce the overhead of processor time, it is proposed to hardly fix threads to physical poisons and refuses to use the HT mode. To control HT technology, each thread uses an API affinity mask, the *SetThreadAffinityMask* function, wherein one logical bit corresponds to each logical processor. To turn off HT on 4 logical processors of the platform, mask 0101b allows the use of one logical processor in each physical core. While, the Operating System (OS) will assign the computational thread to the first logical core, and will assign an idle cycle to the other and prevent migration of threads from busy cores to free ones. In this case, the core will consider that only one input should be executed, and will not switch the pipeline between the two inputs.

To reduce memory access operations outside the processor core, using the static load balancing method [7], pre-determining the part of the DE processed by each processor core. This balancing is effective due to the presence of a priori information about the object. Opposite distribution options are the implementation of all model equations on one or each physical node of the model on a separate processor core. This is determined by the criterion of feasibility in real or accelerated time scale, as well as providing opportunities for complicating the model in the future. The model proposes to distribute the computational usage equally between processor cores: Carriage oscillation equations (1) and (2), oscillations equations of the first trolley(2) and (3), assign the oscillation equations of the second trolley (4) and (5) for execution by separate threads, respectively, Thread 0, Thread 1,and Thread 2.

The distribution criterion also allows the connectedness of the model equations in variables:
- Equable loading of $\tau_i$ cores by threads.
- RT cycle scalability margin.
- Compliance with the structure [19] of the HIL platform.

In the studied model, flow blocks in the RT cycle operate with a working data set. That fits in the cache, and the processor will practically not swap data from memory. In addition, not all read-only access to global variables requires synchronization. Due to the separation of the integration cycles of phase variables into program blocks, there is no simultaneous recording from threads into related variables.

The use of parallel separation in the model with data decomposition provides an increase in the volume of processed data when refining and complicating the model. To do that, established all calls of threads to global variables, assigned all phase variables and their derivatives $Z_k$, $Z_1$, $Z_2$, $Z_1^{'}$, $Z_2^{'}/Z_k$, $\varphi_k$, $\varphi_k^{'}$ as shown in Figure 2.

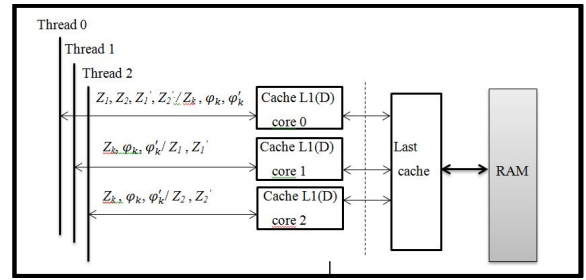

**Figure 2:** Scheme of accessing threads to global variables

The slash (/) symbol separates variables that are written to and read from the cache. The Figure shows that no conflicts are overwriting to the same variable.

### 4.3. Methods of Low-Level Model Optimization.

Low-level optimization of the software model is based mainly on the methods collected in [27]. Focus on optimizing data access, since the code between the transition operations is linear and the processor efficiently preloads the memory.

Access of parallel threads to shared phase variables is protected by synchronization means implemented in hardware or software. Due to the parallelism in the developed model, atomic operations are used that ensure data coherence [33].For modern processors, 32-bit read and write are atomic if the variables are aligned (padding).However, during synchronization, such access is not guaranteed. For 64-bit variables, operations are atomic in 64-bit Windows. Reading/writing to variables of other sizes is not guaranteed to be atomic on any platform [34].The use of atomicity eliminates the possibility of proactive execution and prevents false memory allocation. Also, the advantage of atomic operations is that they are faster compared to locks and are not dependent on deadlocks. When developing a software model, it is decided to align the initialized phase variables at addresses that are multiples of eight (__attribute__ ((aligned (64)))).Data management after reading/writing to the L1d cache is performed by the hardware negotiation protocol.

The compiler and the operating system provide software coherence support methods. When developing a model for preserving RT, it is necessary to take into account the optimizing transformations of a compiler performing software optimization. When executing the model, modern processors with hardware can rearrange the instructions for reading/writing memory to optimize. Command reordering occurs when compiler optimization is enabled. The .NET platform has its memory model, which is independent of the memory model of a particular processor [35].In the .NET memory model permutation is allowed to read and read/write operations. For write-write operations, it is prohibited. However, for the accepted distribution of equations between the cores in this model, this situation does not arise. Most compilers do data alignment on their own, but the strategy they use to create RT models may not be effective. The higher the spatial locality in the model, that provided by the proactive sampling mechanism during execution, the less is the load on the bus for writing data to the cache. Microsoft Visual C++ places the variables in memory in the order of declaration in the program, and the global variables are aligned according to their size. In this case, the compiler will place them close to each other, which can lead to their getting into one cache bank. On Intel processors, an associative data cache is available for reading/writing data if accesses to different cache banks.When placing data in the developed model, it planed so that there are no delays due to simultaneous access to the same cache banks.This ensures that frequently modified data does not fall into the same cache line.In addition, using alignment eliminates the problem of false separation and the phenomenon of "flushing the cache".

Hardware methods solve the coherence problem more efficiently. On Intel processors, each local cache controller contains a bus tracking unit. Monitoring and managing all transactions on the shared bus according to the MESI/MESIF protocols. The protocol defines general rules; specific situations are not specified by it and this is determined by specific implementation, processor model.If the program issues a lock, then global synchronization does not occur, the core flushes the Store buffer to its local cache.In the main memory and caches of adjacent processors, data is not immediately written. Thus, the data exchange between the model threads with shared memory takes into account two operations: direct data transfer between the cores and ordering of this transfer [36].Coherence in the model program may be violated as a result of model recording phase variables, although they occur much less frequently than reading.However, given that write buffers are available not only for writing but also for reading data, using store forwarding [37] by the core reduces the access time to phase variables in the model.Using write buffering when creating a model allows you to defer physical data upload to the L1d cache, but requires that the data be atomic.

For the model under development, a significant part of the data exchanged by the cores is used at one time by one of the threads. The individual phase variables recording of only one of the corresponding threads performs the model.Each quantity and size of the core write buffers, for a given distribution of the model between the cores, a small amount of data for exchange between threads, and sufficient L1d cache for the model under study, allow variables to be stored without cache misses on the RT cycle.Given the continuous and smooth nature of the change in the model phase variables, can assuming that if the result is computed by some processor core and became visible to other cores, then it will be used immediately at the iteration of numerical integration.Thus, thread execution will keep these models internally consistent.

When developing a simulation program, it was decided not to use software prefetching. Since its optimal strategy depends on many factors: the memory type, its latency and access time to it, the clock frequency of the system bus and the core frequency of the processor, cache policies, and cache-line lengths.

## 5. EXPERIMENTS

To determine the optimal distribution of equations over control threads, the memory allocation of a multi-threaded application, it is necessary to determine information about the target system.The identification scheme of logical processors in the Intel architecture is contained in sheets 1, 4, and 11 of the CPUID instruction.

The modeling system based on the Sandy Bridge microarchitecture has 4 cores (see Figure 3a), and a three-level subsystem cachememory (see Figure 3b), defined by these well-known utilities.

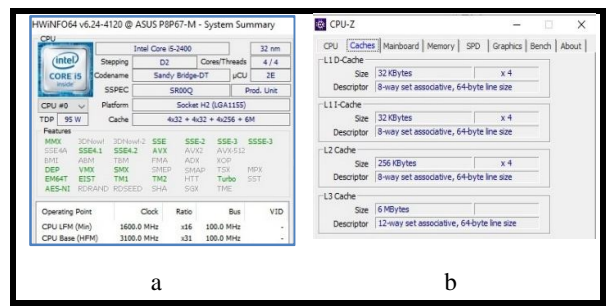

**Figure 3:**Results of the utilities

L1 first-level cache memory has an 8-channel 32-kilobyte data cache (L1d), a 4-channel 32-kilobyte instruction cache (L1i): it is included in the processor core and simultaneously supports two memory read operations (on ports 2 and 3) and one record in memory (on port 4).Each processor core has a unified 8-channel cache of the second level (L2) of 256 KB in size, with two pairs of cores sharing L2. Layer 3 cache (L3 / LLC) is 16-channel, shared by four processor cores, and connected to a ring interconnect.Moreover, the algorithm for filling the cache memory is inclusive (L3 cache - including relations to L1 and L2).The size of the cache line in the processor family does not change and is 64 bytes. All three caches are 8-channel dial-associative.The degree of associativity (8 way) and the size of the cache bank are one of the important characteristics when implementing threads. Physically, the L3 cache is divided into banks, while you can write or read in each bank individually. The entries quantity in L1d is 32kb / 64 = 512 and the number of banks is 512/8 = 64.To avoid thrashing during model implementation, variables should not be located at addresses that are a multiple of the cache bank size.

Additionally, the following necessary data on the target system were obtained from the technical documentation.If there is a competition between the main memory addresses for the same cache lines, the MESI coherence protocol provides a response to the coherent (snoop) request of another Central Processor (CP). When a query crosses a cache line, getting a 5-step penalty.The L1d cache is non-blocking and, after a miss, it can accept further requests during the pre-load data. Processor caches operate by the exchange protocol of the modified version of the QuickPath Interconnect (QPI) bus. Using the forward write mode with buffering and write back.

To determine the RT fulfillment criterion during system development, it is necessary to determine the maximum iteration time for each parallel model thread, in the Worst-Case Execution Time (WCET), i.e.$\tau_i (i = 1,2,...,n)$. Although invariant Time Stamp Counter (TSC) queries allow getting the processor ID simultaneously with the counter value, and having much lower overhead for multi-core processors, Microsoft does not recommend using RDTSCP calls for synchronization with high resolution [38].In addition, the introduction of energy-saving technology in modern processors introduces additional requirements for software measurements, and there are no guarantees that the processor cycles on all the cores of one system will be synchronized.On the instrumental CMC, by the criterion of resolution and accuracy, the hardware used High Precision Event Timer (HPET) of the processor Southbridge with equal access from any core and independent of the core frequency. Before using it in Windows 10, checking HPET via the command line as administrator after entering the command: *bcdedit/set* using platform clock true, and it is accessed in the model using the time-stable *QueryPerformanceCounter* Win32 API functions.When measuring in the developed software model, the overhead of these functions at the beginning and end of the measured code was taken into account.For an instrumental system, the total average performance of two functions is 6,000 ticks of HPET.The measurements used an unprivileged serialization instruction CPUID, forcing the CP to complete each previous code instruction before continuing with the program.This ensures that only the measured code will be executed between calls.The threads single execution (iteration) times of model 1.2 are approximately equal and amounted to 0.018 ms, and for thread 3, the time is 0.017 ms.Thus, the condition for the existence of the cyclic schedule RT is fulfilled with a margin and does not require splitting the threads into blocks.

When developing a model program for a CMC system, the tools that the programmer uses are important. The class of development tools that generate native code in comparison with the generation of bytecode occupies a large share of modeling RT systems. Automatic parallelization tools, when used in optimizing compilers, are more difficult to determine the correctness and efficiency of optimizations compared to a programmer [39].Despite the availability of the best parallelism languages for data management, for the current level of scientific computing, RT uses versions of the C language [40], which less isolate the problematic programmer at all stages of the model from the compiler to the processor.During the research, the model is implemented in C++ in the Visual Studio (VC) environment.TheVS choice is explained by the presence of multithreaded profiling tools based on Spatio-temporal locality.

To synchronize the integration steps, the *WaitForMultipleObjects* function was used, which puts the thread in the standby state and, unlike the *SuspendThread/ResumeThread* functions, is designed specifically for organizing synchronization. When choosing an event as the simplest and most fundamental synchronizing object from two of its types, one is selected whose state is manually controlled by the *SetEvent/ResetEvent* functions.In this case, in comparison with the use of mutexes, there is no effect of flipping the cache.To suspend a thread and put it in a waiting state for an object to be released, use the *WaitForSingleObject* function.The phase variables behavior, as modeling result, based on accepted design decisions, using the Gnuplot utility, is shown in Figure 4.For the above results, when controlling threads by the cyclic schedule, a thread 2 call was made with the computation of $Z_k, \varphi_k$ five times less than others.This ensures the implementation of different inertia in the model.The results obtained coincide with the simulation results by all threads with the same frequency.

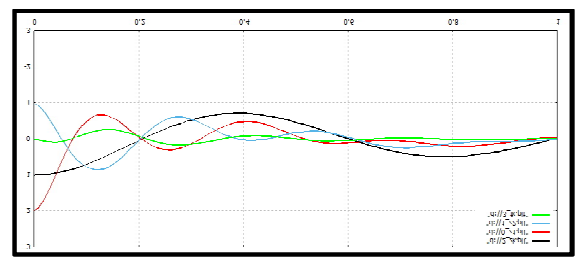

**Figure 4:** The result of modeling a dynamic system

Verifying the model, based on only the summary data of the built-in performance counters, is not enough; it is also necessary to visualize the profiling data [41].To measure the quality of the software model, the authors used the built-in profiler of the development environment After analyzing the trace, Concurrency Visualizer displayed the results on the trace report page, as shown in Figure 5.

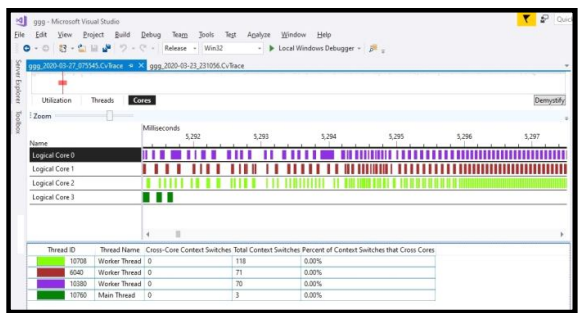

**Figure 5:** The Concurrency Visualizer core view for the model

The "core" representation when displayed, confirms the planned execution of the model threads on the cores of logical processors and the absence of the multi-core migration problem. The result "Legend" is sorted by the context switches quantity of the threads, due to the synchronization of integration steps between the cores in descending order. When migrating threads, context switches between thecores would be more expensive, because cache data is not valid for this thread in another core.If a thread resumes on the core it was running on before, the payload is still in its cache.

To represent "Threads" as shown in Figure 6, the channels for each model thread in the process are presented on the *Y*-axis scale.Logical threads are not filtered and sorted in the order they were created, so the main application thread takes first place. Timelines indicate in green the status of the thread's execution at a given point in time. Using a timeline, an approximately equal working balance is seen between threads that participate in parallel loops.In addition, there is a lack of interference from other processes, which are running in the system and the time it takes to lock synchronization.

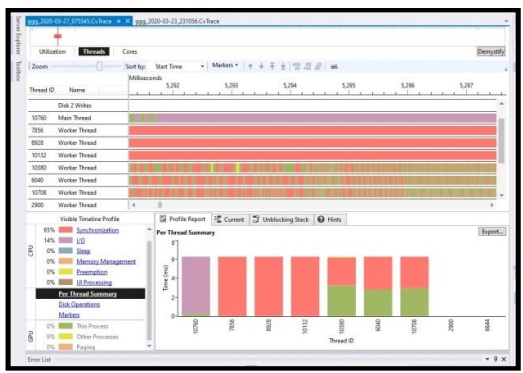

**Figure 6:** The "Threads" view of the Concurrency Visualizer for the model

## 6. CONCLUSIONS

The scientific novelty of the results is the consistent methodology development for optimizing computations. The stages are based on the method proposed by the authors: organization of parallel threads, in a dynamic object complicated model, with various inertia, and differing into consideration of the microarchitecture features of modern computers.

The developed model is a platform for researching both existing and developing rail vehicles.The algorithmic and software tools used in the model that is based on the architecture of CMC systems provide RT mode.Using a programming template in the model with shared memory allocation for child threads working independently of each other is effective.The solution of the problem posed in the article, nonlinear integer optimization, provides a mathematical justification for expanding the criterion for the RT mode existence by dividing the parallel model program flows into blocks.The model reasonably uses cache-friendly access methods. Into consideration, the modeling object features, and providing spatial and temporal locality of the program code during implementation. The practical value of the work lies in the fact that the proposed approach to algorithmic and software-hardware optimization, when building a model with different inertia, is universal.It can be used for the efficient modeling of various dynamic objects. The optimized model will be developed and supplemented via complicating the mathematical description of the object and automating the distribution of parallel implementation threads.

## ACKNOWLEDGEMENT

The authors would like to thank Onaizah Colleges and Donetsk National Technical University for supporting thisresearch.